\documentclass[a4paper, 11pt]{amsart}

\usepackage{amsmath, amssymb, amsthm}
\usepackage[english]{babel}
\usepackage[T1]{fontenc}
\usepackage{graphicx}        
\usepackage{multicol}        
\usepackage[bottom]{footmisc}
\usepackage{url}
\newcommand{\ee}{\mathbf{e}}

\def\e{\epsilon}

\def \be {  \varpi}

\newcommand{\fer}[1]{(\ref{#1})}

\newcommand{\R}{\mathbb R}
\def\be#1\ee{\begin{equation}#1\end{equation}}

\numberwithin{equation}{section}

\newcommand{\bq}{\begin{equation}}
\newcommand{\eq}{\end{equation}}

\theoremstyle{remark}

\theoremstyle{definition}

\newenvironment{equations}{\equation\aligned}{\endaligned\endequation}

\begin{document}

\title[The size distribution of cities: a kinetic explanation]{SIZE DISTRIBUTION OF CITIES: A KINETIC EXPLANATION}

\author{STEFANO GUALANDI}
\address{Department of Mathematics, University of Pavia, 
via Ferrata 1,
Pavia, 27100 Italy}
\email{stefano.gualandi@unipv.it}

\author{GIUSEPPE TOSCANI}
\address{Department of Mathematics, University of Pavia,
via Ferrata 1,
Pavia, 27100 Italy}
\email{giuseppe.toscani@unipv.it}

\maketitle

\begin{abstract}
We present  a kinetic approach to the formation of urban agglomerations  which is based on simple rules of immigration and emigration.  In most cases, the Boltzmann-type kinetic description allows to obtain, within an asymptotic procedure, a Fokker--Planck equation with variable coefficients of diffusion and drift, which describes the evolution in time of some probability density of the city size. It is shown that, in dependence of the microscopic rules of migration, the equilibrium density can follow both a power law for large values of the size variable, which contains as particular case a Zipf's law behavior, and a lognormal law for middle and low values of the size variable. In particular, connections between the value of Pareto index of the power law  at equilibrium and the disposal of the population to emigration are outlined. The theoretical findings are tested with recent data of the populations of Italy and Switzerland.
\end{abstract}

\maketitle

\section{Introduction}

The study of social and economic interactions in multi-agent systems led to an interesting variety of mathematical models which, starting from elementary interaction rules,  have produced various emergent phenomena \cite{CFL, NPT, PT13, SK13}. In socio-economic modeling, these systems are composed not by particles but by humans, and every individual usually interacts with a very limited number of peers, which appears negligible compared to the total number of people in the system. Nevertheless,  the phenomena are characterized by unexpected global behaviors, like the formation of very stable curves for the wealth distribution 
\cite{BST,Ch02, CCM, ChChSt05, CPP, CoPaTo05,DY00, DMT, DMT1, DT08,gup,Ha02,IKR98,MD,MaTo07,Sl04,TBD}, or the emergence of consensus about a specific issue \cite{BN1,BN2,BN3,BeDe,Bou,Bou1,Bou2,BrT15,CDT,Def,DMPW,GGS,GM,Gal,GZ,MG,SW,Tos06}. 

The description of apparently different social phenomena has its common basis in statistical physics. In particular,  methods borrowed from
kinetic theory of rarefied gases have been successfully used to construct master
equations of Boltzmann type, usually referred to as kinetic equations, describing the
time-evolution of the number density of the population and, eventually, the emergence
of universal behaviors through their equilibria  \cite{CaceresToscani2007,NPT,PT13}.

Among the various emergent phenomena which have been studied so far,  one is related to the understanding of the reasons why the distribution of urban agglomeration, above a certain size, is well represented by a power law. This phenomenon is striking, since historically, cities have seen birth, growth, competition, migration, decline and death, but nevertheless their distribution with respect to size appears very uniform across cultures and economies \cite{Bat, Bat1}. While this behavior was first noticed by Auerbach \cite{Auer} one century ago, a convincing social explanation came only half a century later by Zipf's theory \cite{Zipf}, who attributed the statistical formation of a distribution with power tails  to the least effort principle of human behavior. Let us denote by $h(v)dv$ the number of cities having the population size between $v$ and $v+dv$, and by
 \[
R(v) = \int_v^\infty h(w) \, dw
 \] 
the associated cumulative probability distribution function (the rank), which provides the fraction of cities with size greater than $v$. Then in Zipf's law the rank $R(v)$ of a city with population equals $1/v^\gamma$, with $\gamma$ approximately equal to one. By direct inspections
Zipf verified that the scaling exponent $\gamma =1$ was very close to reality for most societies and  across time. 

A theoretical derivation of Zipf's law for cities has been the object of several studies. Gabaix \cite{Ga99} introduced a random growth for cities which in a suitable limit converge to Zipf's law. Zanette and Manrubia \cite{ZM} 
proposed a city formation model based on a  multiplication and diffusion process, and find that their results also reproduce Zipf's law. Other studies obtained the formation of Zipf's law as a direct result of migration \cite{MZ}, by assuming suitable pairwise interactions between inhabitants of different cities. Also  Zipf's law with exponent one was  obtained by Gabaix resorting to simple economics arguments \cite{Ga09}. A slightly different mathematical modeling has been introduced in \cite{GCCC}, where growth and organization of cities has been modelled as a resource utilization problem, where many restaurants compete, as in a game, to attract customers using an iterative learning process. The case of restaurants with uniformly distributed fitness was shown to give rise to Zipf's law for the number of customers.

On the other hand,  variations to this behavior have  been observed for countries like  the former USSR or China countries \cite{Ben, GB}. While the cumulative probability distribution function of these nations follows a power law for large sizes of the cities, the correct exponent $\gamma$ has been shown to take values different from $1$ \cite{Ga16, New}. More recently, the validity of Zipf's law for city sizes has been questioned in a number of papers \cite{Ee04, Ee09, Le09, Mal, Roz,Ioa}. In particular, lognormal distribution was proposed as the correct one for city size \cite{Ee04}, while a recent analysis of the size distribution of US cities \cite{BRS} arrived to the drastic conclusion that the power law behavior does not hold even in the tail.  

Looking for a good fitting of city size distribution in the lower and middle part of the size has been the object of a number of recent papers \cite{Cal,DLDA,GZS,Gom,GRSC, LDD,LD1,LD2,PR,Ram}. The conclusion that can be extracted from these accurate simulations based on real data of large countries (mostly India and Unites States) is that lognormal distribution and its variants provides an accurate description of the middle part (in agreement with the claim of Eechout \cite{Ee04}).  
 
While very accurate in recovering the city size distribution function, in general these contributions do no enter into the mechanism responsible of the formation in time of this behavior, leaving the mathematical modeling of the city size formation largely unexplored. Looking at the pertinent literature, the interested reader can however extract a number of shared reasons.  Among others, it seems established that the main phenomenon leading to the formation of cities is  the tendency of inhabitants to migrate, tendency which relies in both cultural and socio-economic reasons, first and foremost the search for new and better living conditions.  As discussed in \cite{MZ}, this is a relatively recent behavior. In very primitive times a small community (even a family)  was able to perform all necessary activities to survive, and  there was no need to aggregate with other families beyond the size of a tribe. This is no more true in modern times, where mutual cooperation and competition brings people to live together.  Clearly this tendency to aggregate does not work in a single direction, since while a larger part of population prefers to live in a big city, another part prefers to move towards cities of middle size, with the hope to reach a better quality of life.
Note that migration of population across cities can be justified on the basis of other motivations, including the possibility to adjust for resources \cite{BGS, GCCC}. In any case, as it happens in all social phenomena, there is a certain degree of randomness in the process, which takes into account the possibility that part of the variation in size of cities could be generated by unforeseeable reasons.  

In a related socio-economic context, the behavior of wealth distribution in western societies exhibits  many points of contact with the formation of city size. Indeed, the formation of a small class of rich people is related to the common desire of the agents to improve their social condition. This is obtained by trading, which in this latter case plays the role of migration in the former. In the model of
formation of personal wealth introduced in \cite{CoPaTo05},  similarly to binary interactions between particles velocities in the classical kinetic theory of rarefied gases,  the variation law of wealth of agents in a binary trade has been based on two main assumptions: the saving propensity \cite{ChaCha00, Ch02, PCG} and the random risk, main novelty of the modeling in \cite{CoPaTo05}.  
Then, the time behavior of wealth distribution has been fruitfully studied by means of  kinetic models of Boltzmann type, that  largely justified the formation of Pareto tails \cite{Par} on the associated cumulative probability distribution function of wealth on the basis of few universal exchange rules in linear trading interactions \cite{NPT, PT13}. Indeed, as the wealth distribution example learns, one of the main consequences of kinetic modeling of social phenomena is that the microscopic law of variation of the number density consequent to the (fixed-in-time) way of interaction, is able to capture both the time evolution and the steady profile of the density, in presence of some conservation law  \cite{NPT,PT13}. 

Following this line of thought, we will discuss the (eventual) appearance of Zipf's law in the distribution of city size by resorting to fixed microscopic rules describing the evolution of city size in terms of certain citizen's behaviors, that mainly motivates immigration and emigration. 
Starting from these microscopic detailed interactions,  we will construct kinetic models of Boltzmann and Fokker--Planck type, which allow to obtain in some cases the explicit shape of the underlying equilibrium. 

The theoretical findings will be tested with real data of Italian and Swiss population. From the fitting of these data, it it will  appear that the city size distribution is a very composite phenomenon where the behavior of  the population can not be identified in a unique way. A good fitting is indeed obtained by splitting the population in parts which, while homogeneous in their behavior, differ from each other in scope. Surprisingly, the analysis of data from the Italian and the Swiss population show that an extremely good fitting is obtained by resorting to a splitting of the populations in two classes. The behavior of one class determines the formation of the lognormal distribution observed in the middle part of the distribution, while the power law behavior of the population of large cities is due to the second class.   

The kinetic approach used in this paper allows to a better understanding of the possible motivations behind the formation of city size distribution. In particular, it allows to obtain an analytic description of the process of city size forming.  

The forthcoming analysis will be based on the asymptotic relationship between Boltzmann-type equations based on  linear interactions with one-dimensional Fokker--Planck type equations with variable coefficients of diffusion and drift \cite{FPTT17}. Then, the explicit form of the underlying equilibria follows easily by integration of an ordinary differential  equation of first order.

\section{Kinetic modeling of Zipf's law}

Following the basic rules of kinetic theory \cite{PT13}, we will focus on the evolution of a multi-agent system in which agents are identified with urban agglomerations (for simplicity we will denote these entities with the name  cities). Each agent (city) will be entirely characterized by a number  $v $, that will indicate the number of its inhabitants. While it is clear that $v$ is a natural number, to avoid inessential difficulties, we will simply assume in the rest of the paper that $v \in\R_+$.  As briefly discussed in the introduction,  the basic assumption is that the number of inhabitants of a city will essentially grow by immigration, while the size can change for various reasons, and exhibit random fluctuations. This process  can be fully described by a sequence of internal elementary variations of size coupled with immigration of inhabitants from  a fixed environment.  
Hence, each elementary variation  of the size $v$ of a city  is the result of three different contributes (cf. \cite{PT14} for a similar modeling),
 \be\label{lin}
 v^* = v -E(v)v + I_E(v)z + \eta\, v.
 \ee  
In \fer{lin} the variable $z \in \R_+$ indicates the amount of population which can migrate towards a city from the environment. It is usual to assume that this value is sampled by a certain known distribution function, which characterizes the environment itself. 

The functions $E(v)$ and $I_E(v)$ describe the rates of variation of the size $v$ consequent to internal (respectively external) mechanisms. The \emph{internal} rate of variation $E(\cdot)$ is assumed in the form
 \be\label{rata}
 E(v) =  \lambda \frac{\left(v/\bar v\right)^\delta -1 }{\left(v/\bar v\right)^\delta + 1 }.
 \ee
The rate $E(\cdot)$ has been designed to satisfy at better the main aspects linked to migration. The value $\bar v$ in \fer{rata} defines an \emph{ideal city size}.  For Italian cities, the existence of an ideal size depends on well consolidated historical reasons, that  determined in most inhabitants the conviction that the best quality of life can be reached in cities of middle size (the so-called province cities). In \fer{rata}  the constant values $0 \le \lambda < 1$ and $0 < \delta < 1$  quantify the intensity of the rate of migration around the ideal size $\bar v$.  

Note that the rate $E(v)$ can assume both positive and negative values. It is negative when the city size $v$ has a value below the ideal size $\bar v$, and positive in the opposite situation. Hence, this quantity describes the tendency of the population to reach the ideal size $\bar v$ in the case in which $v \not= \bar v$. By construction, independently of $\delta$, $E(\cdot)$ satisfies the bounds
 \be\label{bb}
 -\lambda \le E(v) \le \lambda.
 \ee
Consequently, given the size $v$,  the value $\lambda\, v$ describes the maximal amount of population that can emigrate from the city or migrate towards it in a single microscopic variation.  The presence in \fer{lin} of the minus sign in front of the function $E(v)$ is due to the obvious fact that the population tends to increase when $ v < \bar v$, since people aims in living in a city of population $\bar v$, and to decrease  if $v >\bar v$. 

A further important property is that the rate $E(\cdot)$  is skewed with respect to the value $\bar v$. This follows from the fact that \fer{rata} has to represent at best the main economic aspects  related  to the variation of the population in a city. While  the increasing in size of a city is  linked to a number of positive economic effects, like urbanization and work opportunities, the decreasing in size is accompanied by various problems, including among others the expensive reuse of abandoned areas, as well as the decrease of job opportunities.  Hence, the function $E(\cdot)$ is chosen to possess most of the properties of a value function in the spirit of the pioneering motivations of prospect theory by Kahneman and Twerski \cite{KT, KT1}. This means that, given a certain value $0 < s < 1$
   \[
 - E\left(1-s \right) > E\left(1+s \right).
 \]
This behavior reflects the fact that, given  two cities with size differing of a fixed quantity from the ideal size  $\bar v$ from below and above, a city which has a population below the expected ideal size will have a more pronounced increase of the number of its inhabitants than the decreasing of inhabitants of a city with a size bigger than the ideal one. Note however that the intensity of this variation depends on $\delta$, and it is directly proportional to $\delta$ itself.

Going back to the description of the terms in \fer{lin},  the non negative function $I_E(v)$ provides a measure of the immigration rate, and still quantifies this process in terms of the pre-interaction size number $v$. Finally, the term $\eta\, v$  describes the random fluctuations of the population of the  city. This term takes into account the unpredictable random changes of the size, assumed, without loss of generality, of zero mean and finite variance. Since  $v$ varies  on  the positive half-line,  the random parameter needs to be suitably chosen  to insure that the interaction \fer{lin} maintains the value $v^*$ on the same domain. Taking bound \fer{bb} into account,  a consistent interaction is obtained by assuming that 
 \be\label{posi}
\eta \ge \lambda -1
 \ee
In this case, starting from a city of size $v \ge 0$ it is insured that the interaction \fer{lin} gives $v^* \ge 0$.
 
In reason of the fact that we want to understand the role of the internal migration rate $E(v)$ in the formation of city size distribution, we will start by assuming that the function  quantifying the rate of immigration from the environment towards cities is constant, $I_E(v) = I_E$.  In a second step, we will verify, looking at the resulting kinetic model, that this choice does not change in a significant way the tails of the steady distribution. 

The study of the time-evolution of the distribution of the city size  $v$ consequent to
interactions of type \fer{lin}  can be studied by
resorting to the modeling assumptions of kinetic  theory  \cite{PT13}.  Let us indicate by $f=
f(v,t)$ the density of cities which at time $t >0$ are represented by their
size $v \in \R_+$. Then, the variation in time of $f(v, t)$  due to \fer{lin} obeys to a linear
Boltzmann-like equation. This equation can be fruitfully  written
in weak form. It corresponds to say that the solution $f(v,t)$
satisfies, for all smooth functions $\varphi(v)$ (the observable quantities)
 \begin{equation}
  \label{line-w}
 \frac{d}{dt}\int_{\R_+}f(v,t)\varphi(v)\,dv  = \frac 1\tau
  \Big \langle \int_{\R_+\times \R_+} \left( \varphi(v^*)-\varphi(v) \right) f(v,t)\, \mathcal E(z)
\,dv\,dz \Big \rangle.
 \end{equation}
In \fer{line-w}, the positive constant $\tau$  measures of the frequency of interactions, and plays the role of the Knudsen number in classical kinetic theory \cite{Cer, Cer94}. Also, the symbol $\langle \cdot \rangle$ represents mathematical expectation. Here expectation takes into account the randomness of \fer{lin}, expressed by the  parameter $\eta$. The function $\mathcal E(z)$, $z\in \R_+$ defines the distribution of the size of the environment. It is natural to assume that $\mathcal E(z)$ is a probability density function with finite moments up to some order larger than two. In particular, the finite average value of the environment will be denoted by $M_E$.
 \be\label{mea4}
 \int_{\R_+} z \, \mathcal E(z) \, dz = M_E.
 \ee
 Hence, the quantity $M_E$ measures the mean value of the reservoir from which a city can receive possible immigrants.
The meaning of the kinetic equation \fer{line-w} is clear. At any positive time $t >0$, the variation in time of the distribution of the city size $v$  (the left-hand side) is due to the change in $v$ resulting from interactions of type \fer{lin} between the  city and its environment. This change is measured by the interaction operator at the right-hand side. 

The kinetic equation \fer{line-w} describes the evolution of the density, and allows us to study, at least numerically, the long-time behavior of the sizes of cities, by recovering a macroscopic universal behavior. 

Note that, the choice $\varphi(v) = 1$ in \fer{line-w} shows that the kinetic equation is mass preserving, so that, if $f_0(v)$ denotes the initial probability density functions of city sizes, at each subsequent time
 \be\label{mass}
 \int_{\R_+}f(v,t)\,dv = \int_{\R_+}f_0(v)\,dv = 1, 
 \ee
and the solution to the kinetic equation remains a probability density. 

The choice $\varphi(v) = v$ in \fer{line-w} allows to study the evolution of the mean size value $m(t)$, defined by 
 \be\label{mom1}
m(t) =  \int_{\R_+}v \, f(v,t)\,dv .
 \ee
The evolution of the mean value satisfies the differential equation
 \be\label{me-lin}
 \frac{d}{dt}\int_{\R_+}v \, f(v,t)\,dv = \frac1{\tau}\left( I_E\,M_E - \int_{\R_+}v \, E(v)\, f(v,t)\,dv  \right), 
 \ee
that is not explicitly solvable. If the initial mean value, say $m(0)$,  is bounded, it can be shown that equation  \fer{me-lin} implies the uniform boundness in time of $m(t)$. In fact, since 
 \[
 E(v) = \lambda\left( 1 - 2 \frac 1{(v/\bar v)^\delta  +1} \right),
 \] 
one has the identity
 \[
 - \int_{\R_+}v \, E(v)\, f(v,t)\,dv  = \lambda\, \left( 2  \int_{\R_+} \frac v{(v/\bar v)^\delta  +1} \, f(v,t)\,dv - m(t)\right). 
 \]
Therefore, resorting to the inequality
 \[
 \frac v{(v/\bar v)^\delta  +1} \le \bar v^\delta \, v^{1-\delta},
 \]
 one obtains
 \[
 \int_{\R_+} \frac v{(v/\bar v)^\delta  +1} \, f(v,t)\,dv \le \bar v^\delta \, \int_{\R_+} v^{1-\delta}\, f(v,t)\,dv \le \bar v^\delta\, m(t)^{1-\delta},
 \]
where the last inequality follows from Jensen's inequality. Substituting into \fer{me-lin} one realizes that the mean value satisfies the differential inequality
 \be\label{me-in}
 \frac{d m(t)}{dt} \le \frac1{\tau}\left( I_E\,M_E - \lambda\, m(t) + 2\lambda\, \bar v^\delta m(t)^{1-\delta}  \right).
 \ee
Now, the right-hand side is nonnegative if and only if the positive quantity $m(t) \le \bar m$, where $\bar m$ is the (unique) bounded quantity such that
 \[
I_E\,M_E - \lambda\, \bar m + 2\lambda\, \bar v^\delta \bar m^{1-\delta}. 
 \]
Consequently, if $m(0) < \bar m$, the right-hand side of \fer{me-in} is positive, and $m(t)$ starts to increase, without crossing the value $\bar m$. Conversely, if $m(0) > \bar m$,  $m(t)$ starts to decrease. In any case
 \be\label{me-fin}
 m(t) \le \max \left\{m(0); \bar m \right\}< +\infty.
 \ee
The above computations allow to show that the eventual steady state of the kinetic model has a certain number of moments bounded.
  
However, except in some simple case \cite{BaTo,BaTo2,Sl04},  the Boltzmann-type equation consequent to interactions of type \fer{lin}  does not allow to recover a precise analytic description of the emerging equilibria. A further insight into the large-time behavior of the kinetic equation, and a more accessible description of the possible stationary states can be achieved by resorting to particular asymptotics which lead to Fokker-Planck type equations \cite{CoPaTo05,Tos06}. 

This asymptotic procedure is a well-consolidated technique which is reminiscent of the so-called grazing collision limit, fruitfully applied to the classical Boltzmann equation \cite{Vi,vil2,Vil02} and to the dissipative versions of Kac caricature of a Maxwell gas~\cite{Furioli2012} introduced in \cite{PTo}. For the sake of completeness, we will briefly describe this asymptotic procedure in the next Section. Further details in its general setting can be found in \cite{FPTT17, GT-ec}.

\section{Quasi-invariant size limits and Fokker-Planck equations}\label{quasi}

 As briefly discussed above, among observable quantities, {besides the mass which is  conserved, the first representative ones to be studied are the average values of the density $f$. 
Let $A(v,t)$ denote the deterministic part of the change in size of $v$, as given by \fer{lin}, that is
 \be\label{det}
A(v,z) = I_E \, z -E(v)\, v. 
 \ee
If we set $\varphi(v) = v$ in \fer{line-w} we have
 \[
 \langle v^* - v \rangle =  A(v,z),
 \]
and, consequently
 \begin{equation}
  \label{m-l}
 \frac{d}{dt}\int_{\R_+}v \,f(v,t)\,dv  =  \frac 1 \tau
  \int_{\R_+ \times \R_+} A(v,z) f(v,t)\mathcal E(z)
\,dv\,dz ,
 \end{equation}
 that coincides with  \fer{me-lin}.
Likewise, since
 \[
 \langle {v^*}^2 - v^2 \rangle = \sigma \, v^2 + A^2(v,z) + 2vA(v,z),
 \]
 \begin{equation}
  \label{m-v}
 \frac{d}{dt}\int_{\R_+}v^2 \,f(v,t)\,dv  =  \frac 1\tau
  \int_{\R_+ \times \R_+}\left( \sigma \, v^2 + A^2(v,z) + 2vA(v,z) \right) f(v,t)\mathcal E(z)
\,dv\,dz .
 \end{equation} 
 Note that, proceeding as in the precise evaluation of the mean value performed in the previous Section, it can be proven that the moments at the first two orders of the solution to equation \fer{line-w} remain bounded at any time $t >0$, provided that they are bounded initially.
 
 Let us suppose now that any interaction of type \fer{lin} produces a very small mean change of the size of the city. This can be easily achieved by multiplying the constant $I_E$ and the function $E(\cdot)$ by some value $\e$, with $\e \ll 1$, and the random variable by $\e^{\alpha}$, for some $\alpha >0$. In other words, given a small
parameter $\e$, one considers the scaling
 \be\label{scal}
I_E \to \e I_E,\quad E(\cdot) \to \e E(\cdot), \quad \eta
\to \e^{\alpha} \eta.
 \ee
Concerning the evolution of the average value \fer{m-l} this scaling will produce a small variation in time of the average, given by
\[
 \frac{d}{dt}\int_{\R_+}v \,f(v,t)\,dv  = \frac\e\tau
  \int_{\R_+ \times \R_+} A(v,z) f(v,t)\mathcal E(z)
\,dv\,dz .
 \]
To observe an evolution of the average value independent of $\e,$ it is enough to increase the frequency of interactions. The simplest way to do it is to set $\tau = \e\tau$. Then, if we denote $ f(v,t) = f_\e(v,t)$ to outline the $\e$-dependence,  the evolution of the average value for $f_\e(v,t)$ satisfies
\[
 \frac{d}{dt}\int_{\R_+}v \,f_\e(v,t)\,dv  =  \frac 1\tau
  \int_{\R_+ \times \R_+} A(v,z) f_\e(v,t)\mathcal E(z)
\,dv\,dz ,
 \]
namely the same evolution law for the average value of $f$  given by \fer{m-l}. The reason is clear. If we assume that the interactions are scaled to produce at each time a very small change in the size of cities, to observe an evolution of the average value independent of the smallness of the scaling, we need to suitably increase the frequency of interactions to restore the original evolution. Note that, since the random variable $\eta$ has zero mean, the evolution of the average value does not depend on $\alpha$ in \fer{scal}.

By using the same scaling \fer{scal} into the equation \fer{m-v}, and $\tau \to \e \tau$, the evolution equation for  the second moment of $f_\e(v,t)$ takes the form
\[
\begin{aligned}
 &\frac{d}{dt}\int_{\R_+}v^2 \,f_\e(v,t)\,dv  = \\
  &\frac 1{\e\tau}\int_{\R_+ \times \R_+}\left( \sigma\e^{2\alpha} \, v^2 + \e^2 A^2(v,z) + 2\e\,v\,A(v,z) \right) f_\e(v,t)\mathcal E(z)
\,dv\,dz .
\end{aligned}
 \]  
Hence, by choosing $\alpha = 1/2$, one shows that the evolution of the second moment of $f_\e(v,t)$, for any given $\e \ll 1$  depends on all the quantities appearing in the interaction \fer{lin}, and
 \[
 \frac{d}{dt}\int_{\R_+}v^2 \,f_\e(v,t)\,dv  = \frac 1\tau
  \int_{\R_+ \times \R_+}\left( \sigma v^2 +  2\,v\, A(v,z) \right) f_\e(v,t)\mathcal E(z)
\,dv\,dz  + R_\e(t),
 \]
  where the (small) remainder is given by
 \[
 R_\e(t) =\frac\e\tau \int_{\R_+ \times \R_+} A^2(v,z) f_\e(v,t)\mathcal E(z)
\,dv\,dz.
 \]
If the remainder vanishes as $\e \to 0$, one obtains in the limit a closed form for the evolution of the second moment, given by 
   \begin{equation}
  \label{m-v3}
 \frac{d}{dt}\int_{\R_+}v^2 \,f(v,t)\,dv  = \frac 1\tau
  \int_{\R_+ \times \R_+}\left( \sigma v^2 +  2\,v\, A(v,z) \right) f(v,t)\mathcal E(z)
\,dv\,dz 
 \end{equation}
However, one has to note that, while the scaling \fer{scal} is such that the evolution law of the average value is independent of $\e$, the limit evolution law of the second moment, as given by \fer{m-v3}, is different from the evolution law \fer{m-v}. In particular, for a fixed density $f$ the right-hand side of \fer{m-v} is strictly bigger than the right-hand side of \fer{m-v3}. This shows that the variance of the solution to the kinetic model is strictly bigger than the variance of the (possible) limit density. 
 

The short discussion about moments  contains  the main motivations and the mathematical ingredients that justify the passage from the kinetic model \fer{line-w} to its continuous counterpart given by the Fokker--Planck description. Before to proceed with computations, let us outline that the scaling \fer{scal} is clearly well adapted to the present situation. Indeed, the rates of variation of the population (the function $E(\cdot)$) and the immigration rate (the function $I_E(\cdot)$) in a single interaction are in general negligible with respect to the population size. Also, to observe an appreciable variation of the size one has to wait for an amount of time of the order of years, and this corresponds to increase the importance of the interaction operator to observe it in a unit of time. Clearly, the right balance of scaling to observe the phenomenon in the limit is given by \fer{scal} with $\alpha = 1/2$ and $\tau \to \e\tau$.   

Given a smooth function $\varphi(v)$, let us expand in Taylor series $\varphi(v^*)$ around $\varphi(v)$. Using the scaling \fer{scal} we obtain
 \[
\langle v^* -v \rangle = \e \, A(v,z); \quad  \langle (v^* -v)^2\rangle =  \e^2 \, A^2(v,z) + \e \sigma \, v^2
 \]
Therefore, in terms of powers of $\e$,  we easily obtain the expression
 \[
\langle \varphi(v^*) -\varphi(v) \rangle =  \e \left( \varphi'(v) A(v,z) + \frac 12 \, \varphi''(v)\, \sigma\, v^2 \right) + R_\e (v,z),
 \]
where the remainder term $R_\e$, for a suitable $0\le \theta \le 1$ is given by 
 \be\label{rem}
R_\e(v,z) =  \frac 12 \e^2 \, \varphi''(v) A^2(v,z) + \frac 16  \langle \varphi'''(v+\theta(v^* -v))  (v^* -v)^3\rangle, 
 \ee
and it is vanishing at the order $\e^{3/2}$ as $\e \to 0$. Therefore, since $\tau \to \e \tau$,  the evolution of the (smooth) observable quantity $\varphi(v)$ is given by
\[
\begin{aligned}
 & \frac{d}{dt}\int_{\R_+}\varphi(v) \,f_\e(v,t)\,dv  = \\
 &\frac 1\tau \int_{\R_+\times \R_+} \left( \varphi'(v) A(v,z) + \frac 12 \varphi''(v) \sigma \, v^2 \right) f_\e(v,t)\mathcal E(z)\, dv \, dz + \frac 1\e \mathcal R_\e(t)= \\
 & \frac 1\tau \int_{\R_+} \left( \varphi'(v)( I_E \, M_E -E(v)v) + \frac 12 \varphi''(v) \sigma \, v^2 \right) f_\e(v,t)\, dv + \frac 1\e \mathcal R_\e(t),
 \end{aligned}
 \]
where 
\[
\label{rem3}
\mathcal R_\e(t) = \frac 1\tau \int_{\R_+ \times \R_+} R_\e(v,z)  f_\e(v,t)\mathcal E(z)
\,dv\,dz ,
\]
and $R_\e$ is given by \fer{rem}. Letting $\e \to 0$, and still denoting the limit density by $f(v,t)$ shows that in consequence of the scaling \fer{scal} the weak form of the kinetic model \fer{line-w} is well approximated by the weak form of a linear Fokker--Planck equation (with variable coefficients)
\begin{equations}
  \label{m-13}
 & \frac{d}{dt}\int_{\R_+}\varphi(v) \,f(v,t)\,dv  = \\
  & \frac 1\tau \int_{\R_+} \left( \varphi'(v)( I_E\, M_E -E(v)v) + \frac 12 \varphi''(v) \sigma \, v^2 \right) f(v,t)\, dv. 
 \end{equations}
In fact, { provided the boundary term in $v=0$ produced by the integration by parts vanishes,} equation \fer{m-13} coincides with the weak form of the Fokker--Planck equation
 \begin{equation}\label{FP2}
 \tau \frac{\partial f(v,t)}{\partial t} = \frac \sigma 2 \frac{\partial^2 }{\partial v^2}
 \left(v^2 f(v,t)\right )+ 
 \frac{\partial}{\partial v}\left( (E(v) v - I_E M_E ) f(v,t)\right).
 \end{equation}
{A brief discussion which clarifies  the vanishing of the boundary term can be found in \cite{FPTT17}.}
One of the main advantages in resorting to this asymptotic procedure is that  it is possible to obtain from the Fokker--Planck equation \fer{FP2} its explicit stationary solution, and this follows by solving an ordinary differential equation of first order. 

\section{The stationary distribution}\label{steady}
The stationary distribution of the Fokker--Planck equation \fer{FP2} is easily found by solving the differential equation 
 \be\label{sd}
 \frac \sigma 2 \frac{d }{dv}
 \left(v^2 f\right )+ 
  (E(v) v - I_E M_E ) f =0.
 \ee
Solving \fer{sd} with respect to $g = v^2 f$ by separation of variables gives as unique solution to \fer{sd} the function
 \be\label{equilibrio}
f_\infty(v) = \frac\kappa{v^{2}} \left(\frac v {(1+(v/\bar v)^\delta)^{2/\delta}} \right)^{2\lambda/\sigma}
\exp\left(-\frac{2 I_E\,M_E}{\sigma \, v}\right).
 \ee 
In \fer{equilibrio} the positive constant $\kappa$ has been chosen to fix the mass of the equilibrium density equal to unity. The equilibrium distribution \fer{equilibrio} has a polynomial rate of decay at infinity given by 
 \be\label{dede}
1+  \gamma = 2\left(1 + \frac\lambda\sigma \right).
 \ee
This rate is related to the both the parameters $\lambda$ and $\sigma$ denoting respectively the asymptotic value of the internal rate of migration and the variance of the random fluctuations. It is remarkable that the decay rate does not depend on the values of the ideal size $\bar v$ and of the constant $\delta$. The expected rank $R(v)$ for large values of $v$ in this situation is
 \be\label{rank1}
 R(v) \cong v^{-\gamma} = v^{- 1 -2\lambda/\sigma}.
 \ee
The value $\gamma = 1$, namely the classical Zipf's law  is obtained only for $\lambda =0$ (or $\delta = 0$), namely in absence of the value function $E(v)$ and in presence of a constant rate of immigration from the background. Hence, the limit value $\gamma = 1$ does not correspond to a realistic situation, in that all the economic reasons behind the formation of city size are not taken into account. Also, values close to $1$ for $\gamma$ can be assumed or for small values of the parameter $\lambda$, or for large values of the variance $\sigma$, that is in presence of large diffusion. Since large diffusion is not realistic in most situations, values close to one are assumed for small values of the parameter $\lambda$. Consequently closeness to Zipf's law is typical of countries in which immigration dominates the internal rate of change of city size, or the population manifests only a limited attractiveness towards towns of a certain ideal  size.

We remark that the description of the city size evolution density by means of the Fokker--Planck equation \fer{FP2} allows to obtain the steady state profile for the whole range of the size, thus giving a precise description of the distribution at equilibrium of the population also in cities of medium and small size. There, a precise fitting of the real population of a country (when available) allows to better verify whether or not the proposed model is good enough. 

Concerning the rate of decay for large values of the city size, the profile of the steady distribution does not change in a relevant way by assuming the rate of immigration $I_E(v)$ depending on the size $v$. A natural choice, which generalizes the previous assumption (constant rate) is given by 
 \be\label{good}
 I_E(v) = \mu\frac{v^\alpha}{1+v^\alpha},
 \ee
where $\mu$ and $0< \alpha \le 1$ are positive constants denoting the intensity of the immigration rate. This choice corresponds to fix the rate of immigration from the surrounding increasing with respect to the city size, while tending to a positive constant value as $v \to +\infty$. Hence the rate \fer{good} represents the normal situation of a country in which cities of big size are more attracting than cities of small size, for example in view of the different number of opportunities they offer. 

The stationary distribution of the Fokker--Planck equation \fer{FP2} now solves the differential equation 
 \be\label{sd2}
 \frac \sigma 2 \frac{d }{dv}
 \left(v^2 f\right )+ 
  (E(v) v - \mu\frac{v^\alpha}{1+v^\alpha} M_E ) f =0.
 \ee
Solving \fer{sd2} with respect to $g = v^2 f$ by separation of variables gives as unique solution to \fer{sd2} the function
 \be\label{equilibrio2}
f_\infty(v) = \frac\kappa{v^{2}} \left(\frac v {(1+(v/\bar v)^\delta)^{2/\delta}} \right)^{2\lambda/\sigma}
\, \left( \frac v{(1+v^\alpha)^{1/\alpha}}\right)^{2\mu\,M_E/\sigma}
 \ee 
 By assuming 
  \be\label{con3}
  \frac 1\sigma\left( \mu M_E + \lambda\right) >1,
  \ee
it follows that the equilibrium density decays to zero as  $v\to 0$. Moreover, the equilibrium distribution \fer{equilibrio2} has a polynomial rate of decay at infinity given by \fer{dede}. Hence, the immigration rate \fer{good}, provided $\mu$ is such that \fer{con3} is satisfied, gives the same decay at zero and  the  same tail at infinity of the steady distribution \fer{equilibrio}. 

\section{The appearing of lognormal distribution}

As illustrated in Section \ref{steady}, in our kinetic model Zipf's law is determined by an immigration dominated regime.
In this Section we will discuss the opposite situation, namely a society in which the movements among the inhabitants of the cities are dominant with respect to immigration from outside. 

If we consider as before that immigration from the environment is mainly directed towards cities of big size, this behavior is particularly well-adapted to describe the distribution of cities of small and mean size. Proceeding as in \cite{GT17}, for a given small $\delta$ we rewrite the internal rate of variation as
 \be\label{log}
E(v) =  \delta \lambda \frac 1\delta \frac{\left(v/\bar v\right)^\delta -1 }{\left(v/\bar v\right)^\delta + 1 } 
 \ee
Consequently, if we assume that the variation of size due to the internal rule is small ($\delta \approx \e$), we can consistently write, for $v \le V< +\infty$
\cite{GT17}
\be\label{log2}
E(v) \approx   \e \frac\lambda{2} \log \frac v{\bar v}.
 \ee
Hence, by assuming a dominated internal rate of migration, and a random fluctuation as in Section \ref{quasi}, we can apply to the kinetic equation  \fer{line-w} the scaling 
 \be\label{scal2}
 I_E(v) \to \e^{1+\beta}I_E(v), \quad E(v) \to \e \frac\lambda{2} \log \frac v{\bar v},  \quad \eta \to \e^{1/2}\eta,
 \ee
where now the positive exponent $\beta$ enhances the irrelevant effects of the immigration towards cities of small and middle size. Proceeding as in Section \ref{quasi} (cf. also the computations in \cite{GT17}) we obtain that the limit density, say $g= g(v,t)$ now solves the Fokker--Planck equation  
 \begin{equation}\label{FP3}
 \frac{\partial g(v,t)}{\partial t} = \frac \sigma 2 \frac{\partial^2 }{\partial v^2}
 \left(v^2 g(v,t)\right )+ \frac \lambda 2
 \frac{\partial}{\partial v}\left(  v\, \log \frac v{\bar v}\, g(v,t)\right).
 \end{equation}
Equation \fer{FP3} describes the evolution of the density $g(w,t)$ of the size of cities when the internal rate of change is dominant with respect to immigration from the environment. At difference with the solution \fer{equilibrio} of the Fokker--Planck \fer{sd}, the steady state density does not have fat tails at infinity, and results to be a lognormal density, with parameters linked to the details of the microscopic internal size variation.

The unique steady solution to \fer{FP3} of unit mass is given by the lognormal density
 \be\label{equilibrio2}
g_\infty(v) = \frac 1{\sqrt{2\pi \nu}\, v} 
\exp\left\{ - \frac{(\log v - \mu)^2}{2 \nu}\right\},
 \ee 
where 
 \be\label{para}
 \nu = \frac \sigma\lambda,  \quad \mu = \log \bar v - \nu.
 \ee
 Hence, the  mean and variance of the equilibrium distribution \fer{equilibrio2} are given respectively by
 \be\label{moments}
 m(g_\infty) = \bar v e^{-\nu/2}, \quad Var(g_\infty) = \bar v^2 \left( 1 - e^{-\nu}\right).
 \ee
Note that the moments are expressed in terms of the parameters $\bar v$, $\sigma$ and $\lambda$ denoting respectively the optimal size $\bar v$, the variance $\sigma$ of the random fluctuations and the asymptotic size $\lambda$ of the internal rate function $E$. 
The hypothesis which leads to the appearance of the lognormal distribution of city size, while well motivated by prospect theory, and accurate for cities of middle size, does not take into account the role previously played by immigration. 

This effect can be easily introduced by splitting the population in two parts, the first part characterized by its tendency to move towards cities of middle size (characterized by an ideal size $\bar v_M$), the second one with tendency to move towards cities of large size (characterized by an ideal size $\bar v_L \gg \bar v_M$).  By assuming that the evolution of the distributions $u_1(v,t)$  and $u_2(v,t)$ of the two parts of inhabitants is described by a Fokker--Planck equation of type \fer{FP3}, each one with its characteristic parameters, the distribution of the whole population 
 \[
 u(v,t) = \lambda_1 u_1(v,t) + \lambda_2u_2(v,t), \quad \lambda_1 + \lambda_2 = 1,
 \]
will converge to a steady state given by the weighted sum of two lognormal densities of type \fer{equilibrio2}. Numerical experiments, reported in the forthcoming Section \ref{nume}, show that one can obtain in this way a very accurate fitting of the populations of Italy and Switzerland, and that the parameter $\lambda_1$ that represents the percentage of the population which aims in living in cities of middle size is dominant with respect to $\lambda_2$. 

\section{Numerical Experiments}\label{nume}

In this section, we report our numerical experiments using the open data published by the Italian National Institute of Statistics 
\footnote{http://www.istat.it, last visited June, 20th, 2018.}
and by the Swiss Federal Statistical Office
\footnote{http://www.bfs.admin.ch, last visited June, 20th, 2018.}.
The first dataset contains the size distribution of $8\,006$ Italian cities, ranging from the smallest village, to the largest city, 
which is Rome with $2\,873\,494$ citizens. These data refer to the last official Italian census of 2016.
The second dataset contains the size distribution of $2\,289$ Swiss cities, with the largest city, which is Zurich with $396\,955$ citizens. These data refer to the last official Swiss census of 2014.
Table \ref{tab:summary} reports the basic statistics of the two datasets, giving in order the minimum, the first quartile,
the median, the mean, the third quartile and the maximum values of city size.
Clearly, the basic statistics are clueless about the real distribution of cities size.

\begin{table}[th!]
\centering
\caption{Basic statistics of Italian and Swiss distribution of cities size.\label{tab:summary}}
\begin{tabular}{lrrrrrr}
            & Min & 1st Quart. & Median & Mean & 3rd Quart. & Max \\
\hline
Italy       & 30  & $1\,019$     & $2\,452$ & $7\,571$ & $6\,218$ & $2\,873\,494$ \\
Switzerland & 13  & 642        & $1\,425$  & $3\,638$ & $3\,513$       & $396\,955$ \\
\end{tabular}
\end{table}

Dataset on distribution of cities size are in the literature studied and fitted using a Zipf's law.
However, if we just take the logarithm of every city size and we plot the resulting distribution,
we get what looks like a classical Gaussian distribution, as shown in Figures \ref{fig:1}(a) and \ref{fig:2}(a),
respectively for the Italian and Swiss dataset (and it is almost impossible to distinguish the shape of the two distributions).
Even if we looks to the inverse cumulative functions, given by Figures \ref{fig:2}(b) and \ref{fig:2}(b), it is pretty hard
to distinguish the resulting function from a Gaussian cumulative function.
However, if we analyze the inverse cumulative functions resorting to bi-logarithm plots, it is possible to notice
that a single Gaussian does not capture the trend of the tails of the distribution,
as shown with the red lines in Figures \ref{fig:1}(c), \ref{fig:1}(d), \ref{fig:2}(c), and \ref{fig:2}(d).
We remark that anyway even a single Gaussian is able to perfectly fit the lower tails, which are never captured by the celebrated Zipf's law.

In order to improve the fitting of the distributions also on the higher tails, we have fitted the
distributions of cities sizes using a multi-modal Gaussian model, using the {\tt mixtools} software package
\footnote{https://cran.r-project.org/web/packages/mixtools, last visited, June, 20th, 2018.} 
available in the R statistical programming language. For full details on the {\tt mixtools} package we refer the interested reader to \cite{mixtools}.
Basically, using {\tt mixtools} we were able to fit in logarithmic scale the (steady) distribution of cities size with the mixture of two Gaussian 
\begin{equation}\label{eq:bimodal}
N_\infty(v) = \lambda_1 N(v; \mu_1, \sigma_1) +  \lambda_2 N(v; \mu_2, \sigma_2)
\end{equation}
Table \ref{tab:bimodal} reports the parameters fitted by {\tt mixtools} for both datasets and Figures \ref{fig:3} and \ref{fig:4} shows
the respective probability density functions. It is evident that for both datasets there is a ``dominating'' Gaussian. In fact we have $\lambda_1 = 0.945$ for Italy and $\lambda_1 = 0.967$ for Switzerland.
In both cases, the correction that serves to capture the behavior of the higher tails,
and which has both larger means and larger deviations, is obtained by adding a tiny Gaussian (note the small values of $\lambda_2$). We remark that the blue solid line on top of the histograms represents the corresponding bimodal distributions.
Finally, by looking at the green lines in Figures \ref{fig:1}(c), \ref{fig:1}(d), \ref{fig:2}(c), and \ref{fig:2}(d),
the goodness of fitting cities size distributions with a mixture of two Gaussian is striking evident.

\begin{table}[th!]
\centering
\caption{Mixture of two Gaussians: Model Parameters.\label{tab:bimodal}}
\begin{tabular}{lrrrrrr}
            & $\lambda_1$ & $\mu_1$ & $\sigma_1$ & $\lambda_2$ & $\mu_2$ & $\sigma_2$ \\
\hline
Italy       & 0.945 & 3.371 & 0.563 & 0.054 & 3.993 & 0.731 \\
Switzerland & 0.967 & 3.162 & 0.533 & 0.032 & 3.483 & 0.896 \\
\end{tabular}
\end{table}

\begin{figure}[th!]
\centering
\includegraphics[width=\textwidth]{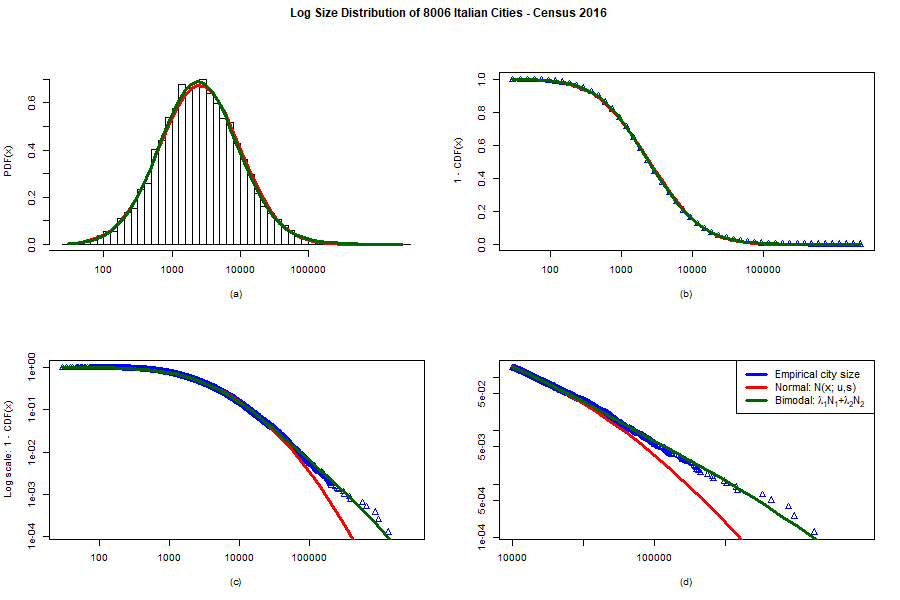}
\caption{Probability distribution function and inverse cumulative functions of Italian cities.}
\label{fig:1}
\end{figure}

\begin{figure}[th!]
\centerline{\includegraphics[width=\textwidth]{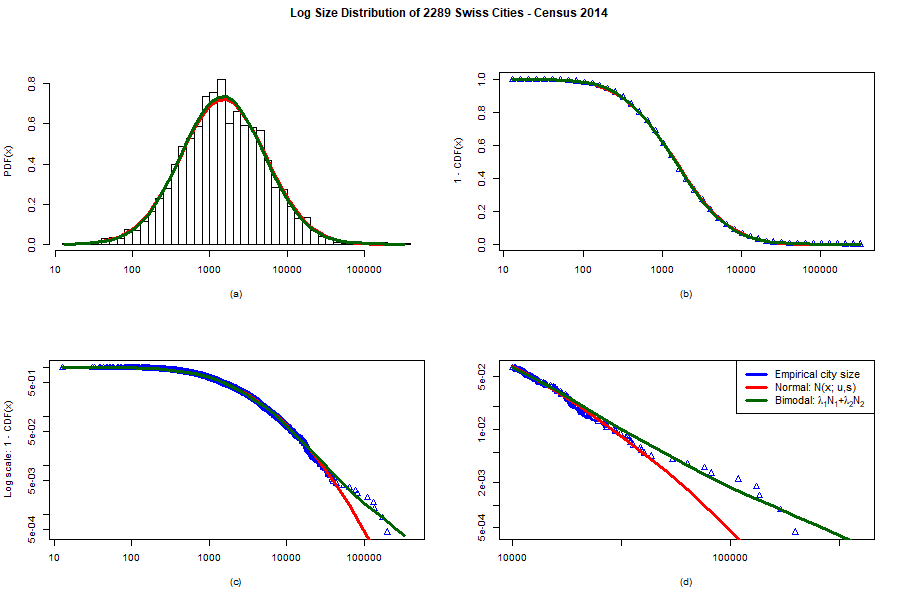}}
    \caption{Probability distribution function and inverse cumulative functions of Swiss cities.}
    \label{fig:2}
\end{figure}

\begin{figure}[th!]
\centering
\includegraphics[width=0.65\textwidth]{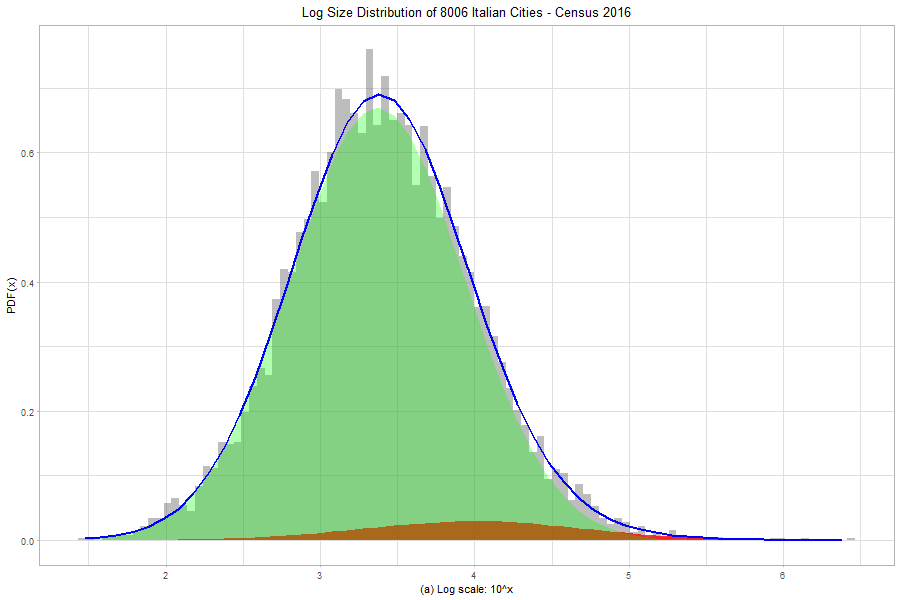}
\caption{Log Size Distribution of 8006 Italian Cities, Census 2016.} \label{fig:3}
\end{figure}

\begin{figure}[th!]
\centerline{\includegraphics[width=0.65\textwidth]{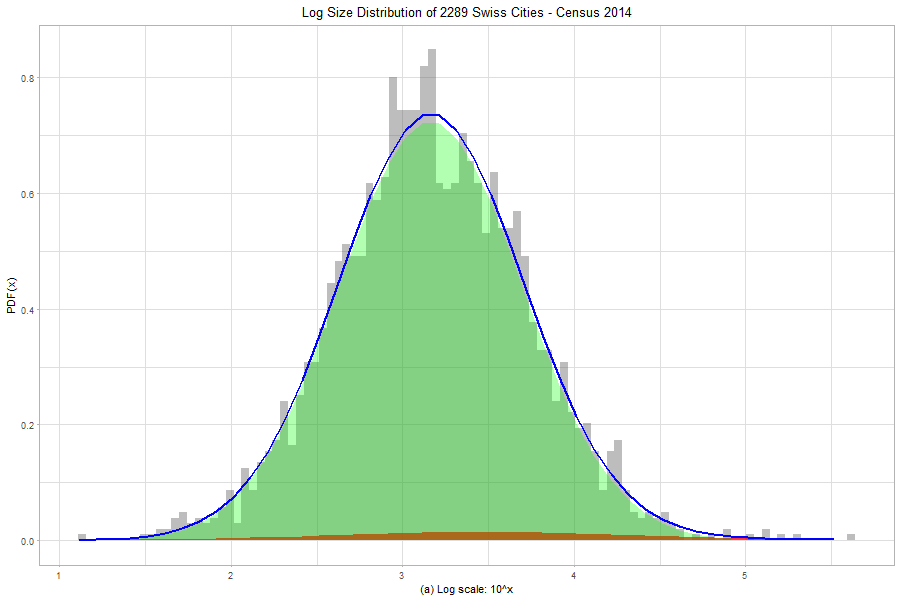}}
    \caption{Log Size Distribution of 2289 Swiss Cities, Census 2014.}
    \label{fig:4}
\end{figure}

\section{Conclusions}

Kinetic modeling is a powerful tool which allows to study a number of collective phenomena in a multi-agent system on the basis of few microscopic rules of interaction. We applied these techniques to the study of the city size in a country, aiming in understanding the elementary mechanism able to produce in time the observed behavior of the distribution of urban agglomerations. The main finding in our analysis is that a power law for large size of city size is obtained in presence of both internal movements of inhabitants between cities and immigration from an external background. Internal movements have been realistically described in terms of a value function in the spirit of the prospect theory by Kahneman and Twersky  \cite{KT, KT1}. 

When the internal movements are dominant, it is shown that the steady city size distribution takes the form of a lognormal density, which perfectly fits the distribution of cities of small and middle size, while a mixture of two lognormal densities is able to carefully describe the whole range of the population. Numerical fitting of data relative to cities of Italy and Switzerland illustrate the goodness of the proposed kinetic modeling.

The present analysis illustrates the difficulties present in identifying in a precise way the city size distribution, and the eventual formation of a power law. At difference with previous attempts, we believe that the kinetic description is helpful to connect in a clear way the microscopic mechanism responsible of the variation of the population size with its global behavior.


\section*{Acknowledgement} This work has been written within the
activities of GNFM and GNAMPA groups  of INdAM (National Institute of
High Mathematics), and partially supported by  MIUR project ``Calculus of Variations''.

This research was partially supported by the Italian Ministry of Education, University and Research (MIUR) through the 
``Dipartimenti di Eccellenza Program (2018--2022)'' - Dept. of Mathematics ``F. Casorati'', University of Pavia. 



\vskip 3cm
\begin{thebibliography}{00}

\bibitem{Ait}
J. Aitchison and J.A.C. Brown, \emph{The Log-normal Distribution},
Cambridge University Press, Cambridge, UK 1957. 

\bibitem{Auer}
F.  Auerbach, Das Gesetz der Bev\"olkerungskonzentration. {Petermanns Geogr. Mitteilung.} \textbf{59}   (1913) 74--76
\bibitem{BaTo}
F.~Bassetti and G.~Toscani.
\newblock Explicit equilibria in a kinetic model of gambling.
\newblock {\em Phys. Rev. E } {\bf 81}  (2010) 066115.

\bibitem{BaTo2}
F.~Bassetti and G.~Toscani.
\newblock Explicit equilibria in bilinear kinetic models for socio-economic
  interactions.
\newblock {\em ESAIM: Proc. and Surveys}  {\bf 47}  (2014) 1--16.


\bibitem{Bat}
M. Batty, Rank clocks. \emph{Nature} \textbf{444} (2006) 592--596.

\bibitem{Bat1}
M. Batty, The size, scale, and shape of cities.
\emph{Science} \textbf{319}, (2008) 769--771.

\bibitem{BGS}
P. Beaudry, D.A. Green and B.M. Sand, Spatial equilibrium with unemployment and wage bargaining: Theory and estimation. \emph{J. Urban Econ.} \textbf{79} (2014) 2--19.

\bibitem{BRS}
M. Bee, M. Riccaboni and S. Schiavo, The size distribution of US cities: not Pareto, even in the tail. \emph{Economics Letters}
\textbf{120} (2013) 232--237.

\bibitem{mixtools}
T.~Benaglia, D.~Chauveau, D.~Hunter, and D.~Young,  mixtools: An R package for analyzing finite mixture models,
  \emph{Journal of Statistical Software}
  \textbf{32(6)} (2009) 1--29.

\bibitem{Ben}
L. Benguigui and E. Blumenfeld-Lieberthal, Beyond the power law--a new approach to analyze city size distributions. \emph{Comput. Environ. and Urban Systems}
\textbf{31} (2007) 648--666.


\bibitem{BN1}
E. Ben-Naim, Opinion dynamics: rise and fall of political parties. \emph{Europhys. Lett.} \textbf{69}, (2005) 671--677.

\bibitem{BN2}
 E. Ben-Naim, P. L. Krapivski and S. Redner, Bifurcations and patterns in compromise processes. \emph{Physica D} \textbf{183},  (2003) 190--204.

\bibitem{BN3}
E. Ben-Naim, P. L. Krapivski, R. Vazquez and S. Redner, Unity and discord in opinion dynamics. \emph{Physica A} \textbf{330},
 (2003) 99--106.



\bibitem{BeDe}
M.L.  Bertotti and  M. Delitala,  On a discrete generalized kinetic approach for mo\-del\-ling persuader's influence in opinion formation processes.  \emph{Math. Comp. Model.} \textbf{48}, 
(2008) 1107--1121.


\bibitem{BST}
M. Bisi, G. Spiga and G. Toscani, Kinetic models of conservative economies with wealth redistribution. \emph{Commun. Math. Sci.} \textbf{7} (4)
901--916 (2009)

\bibitem{Bou}
L. Boudin and F. Salvarani,  The quasi-invariant limit for a kinetic model of sociological collective behavior. \emph{Kinetic Rel. Mod.} \textbf{2}, (2009) 433--449.

\bibitem{Bou1}
L. Boudin and F. Salvarani, A kinetic approach to the study of opinion formation. \emph{ESAIM: Math. Mod. Num. Anal.} \textbf{43}, 
(2009)  507--522.

\bibitem{Bou2}
L. Boudin, A. Mercier and  F. Salvarani, Conciliatory and contradictory dynamics in opinion formation. \emph{Physica A} \textbf{391}, (2012) 5672Â--5684.

\bibitem{BrT15}
C. Brugna and G. Toscani, Kinetic models of opinion formation in the presence of personal conviction, \emph{Phys. Rev. E} \textbf{92},   (2015) 052818.


\bibitem{CaceresToscani2007}
M.J. C{\'a}ceres and G.~Toscani.
\newblock Kinetic approach to long time behavior of linearized fast diffusion
  equations.
\newblock {\em J. Stat. Phys.} {\bf 128} (2007) 883--925.

\bibitem{Cal}
E. Calderín-Ojeda,
The distribution of all French communes: A composite parametric approach.
\emph{Physica A} \textbf{450} (2016) 385--394.

 
 \bibitem{CFL}
C. Castellano, S. Fortunato and V. Loreto, Statistical physics of social dynamics. \emph{Rev. Mod. Phys.} \textbf{81}
(2009) 591--646. 

\bibitem{Cer}
C. Cercignani,
\newblock {\emph{The Boltzmann equation and its applications}}.
\newblock  Springer Series in Applied Mathematical Sciences,
  Vol.\textbf{67}, Springer--Verlag, New York 1988
  
\bibitem{Cer94}
C. Cercignani C., R. Illner and M. Pulvirenti, 
\emph{The mathematical theory of dilute gases}.
  {Springer Series in Applied Mathematical Sciences},
  Vol.\textbf{  106},  Springer--Verlag, New York 1994  

\bibitem{Ch02}
A.~Chakraborti, {Distributions of money in models of market economy}. {\em Int. J. Modern Phys. C} \textbf{13},  (2002)
1315--1321.

\bibitem{ChaCha00} A.~Chakraborti and B.K.~Chakrabarti, Statistical
  Mechanics of Money: Effects of Saving Propensity. {\em Eur. Phys. J. B}
  \textbf{17},  (2000)  167--170.
  
  
\bibitem{CCM} A. Chatterjee, B.K. Chakrabarti and S.S. Manna,
   Pareto law in a kinetic model of market with random saving propensity.
  \emph{Physica A\/} {\bf 335}  (2004), 155--163.

\bibitem{ChChSt05} A.~Chatterjee, B.K.~Chakrabarti and
  R.B.~Stinchcombe, Master equation for a kinetic model of trading
  market and its analytic solution. {\em Phys. Rev. E} \textbf{72},   (2005) 026126.


  \bibitem{CDT}
V. Comincioli, L. Della Croce and G. Toscani,  A Boltzmann-like equation for choice formation. \emph{Kinetic Rel. Mod.} \textbf{2},  (2009) 135--149.

\bibitem{CPP}
S. Cordier, L. Pareschi and C. Piatecki, Mesoscopic modelling of financial markets. \emph{J. Stat. Phys.} \textbf{134} (1),  (2009) 161--184.

\bibitem{CoPaTo05} S.~Cordier, L.~Pareschi and G.~Toscani, On a kinetic
  model for a simple market economy. {\em J. Stat. Phys.} \textbf{120},  (2005)  253-277.
  
\bibitem{Cro}
E.L. Crow and K. Shimizu eds., \emph{Log-normal distributions: theory and application},
Marcel Dekker, New York NY 1988.

\bibitem{Def}
G. Deffuant, F. Amblard, G. Weisbuch and T. Faure, How can extremism prevail? A study based on the relative agreement interaction model. \emph{J. Art. Soc. Soc. Sim.}
\textbf{5},  (2002) 1.

\bibitem{DLDA}
S. Devadoss, J. Luckstead, D. Danfort and S. Akhundjanov,
The power law distribution for lower tail cities in India.
\emph{Physica A} \textbf{442} (2016) 193--196.

\bibitem{DY00}
A.~Dr\v{a}gulescu and V.M.~Yakovenko, {Statistical mechanics of money}. {\em Eur. Phys. Jour. B} \textbf{17},  (2000)
723-729.

\bibitem{DMT}
B.~D{\"u}ring, D.~Matthes and  G.~Toscani, Kinetic Equations modelling Wealth Redistribution: A comparison of Approaches. \emph{Phys. Rev. E},
\textbf{78},   (2008)  056103.

\bibitem{DMT1}
B.~D{\"u}ring, D.~Matthes and G.~Toscani, A Boltzmann-type approach to the formation of wealth distribution curves.  (Notes of the Porto Ercole
School, June 2008)   \emph{Riv. Mat. Univ. Parma} (1) \textbf{8}   (2009) 199--261.

\bibitem{DT08}
B. D\"uring  and G. Toscani, International and domestic
  trading  and wealth distribution.
  \emph{Commun.\ Math.\ Sci.\/}, \textbf{6} (4)  (2008) 1043--1058.

\bibitem{DMPW}
B.~D{\"u}ring, P.A. Markowich, J-F. Pietschmann and M-T. Wolfram.
\newblock Boltzmann and {F}okker-{P}lanck equations modelling opinion formation
  in the presence of strong leaders.
\newblock {\em Proc. R. Soc. Lond. Ser. A Math. Phys. Eng. Sci.},
  {\bf 465} (2009)  3687--3708.
  

\bibitem{Ee04}
J. Eeckhout, Gibrat's law for (all) cities. \emph{Am. Econ. Rev.} \textbf{94}, (2004)
1429--1451.

\bibitem{Ee09}
J. Eeckhout, Gibrat's law for (all) cities: reply. \emph{Am. Econ. Rev.} \textbf{99}, (2009)
1676--1683.

\bibitem{Furioli2012}
G.~Furioli, A.~Pulvirenti, E.~Terraneo and G.~Toscani.
\newblock The grazing collision limit of the inelastic {K}ac model around a
  {L}\'evy-type equilibrium.
\newblock {\em SIAM J. Math. Anal.}  {\bf 44}  (2012) 827--850.

\bibitem{FPTT17}
G.~Furioli, A.~Pulvirenti, E.~Terraneo and G.~Toscani.
\newblock Fokker--Planck equations in the modelling of socio-economic phenomena, \emph{Math. Mod. Meth. Appl. Scie.} \textbf{27} (1)  (2017) 115--158.

\bibitem{Ga99}
X. Gabaix, Zipf's law for cities: an explanation. \emph{Quart. J. Econom.} \textbf{114}, (1999) 739--767.

\bibitem{Ga09}
X. Gabaix, Power Laws in Economics and Finance, \emph{Am. Econ. Rev.}
\textbf{89} (2009) 255--294. 

\bibitem{Ga16}
X. Gabaix, Power laws in economics: an introduction. \emph{The Journal of Economic Perspectives} \textbf{30}, (2016) 185--205. 


\bibitem{GGS}
{S. Galam, Y. Gefen and Y. Shapir}, Sociophysics: A new approach of sociological collective behavior. I. Mean-behavior description of a strike {\em J. Math. Sociology} \textbf{9}, (1982) 1--13.

\bibitem{GM}
S. Galam and S. Moscovici, Towards a theory of collective phenomena: consensus and attitude changes in groups. \emph{Euro. J. Social Psychology} \textbf{21}, (1991) 49--74.

\bibitem{Gal}
S. Galam, Rational group decision making: A random field Ising model at $T= 0$. \emph{Physica A} \textbf{238},  (1997) 66--80.

\bibitem{GZ}
S. Galam and J.D.  Zucker, From individual choice to group decision-making. \emph{Physica A}  \textbf{287}, (2000) 644--659.

\bibitem{GB}
K. Gangopadhyay and B. Basu, City size distributions for India and China.
\emph{Physica A} \textbf{388}, (2009) 2682--2688.

\bibitem{GCCC}
A. Ghosh, A. Chatterjee, A.S. Chakrabarti and B.K. Chakrabarti,
Zipf's law in city size from a resource utilization model.
\emph{Phys. Rev. E} \textbf{90}, (2014) 042815

\bibitem{GZS}
K. Giesen, A. Zimmermann and J. Suedekum,
The size distribution across all cities-Double Pareto lognormal strikes, \emph{Journal of Urban Economics} \textbf{68} (2010) 129-137.

\bibitem{Gom}
E. G\'omez-D\'eniz and E. Calderín-Ojeda,
On the use of the pareto arctan distribution for describing city size in Australia and New Zealand.
\emph{Physica A} \textbf{436} (2015)  821--832.

\bibitem{GRSC}
R. Gonz\'alez--Val, A. Ramos,  F. Sanz--Gracia and   M. Vera--Cabello,
Size distributions for all cities: Which one is best?, \emph{Papers in Regional Sciences} 
\textbf{94}  (2015) 177--196.

\bibitem{GT-ec}
S. Gualandi and G. Toscani, Pareto tails in socio-economic phenomena: a kinetic description. \emph{Economics: The Open-Access, Open-Assessment E-Journal}, 12 (2018-31): 1--17. 
 
 \bibitem{GT17}
S. Gualandi and G. Toscani, Call center service times are lognormal. A Fokker--Planck description. \emph{Math. Mod. Meth. Appl. Scie.}  \textbf{28} (8) (2018) 1513--1527.

\bibitem{gup}
A.K. Gupta, Models of wealth distributions: a perspective. In \emph{Econophysics and sociophysics: trends and perspectives}
 B.K. Chakrabarti, A. Chakraborti, A. Chatterjee (Eds.) Wiley VHC, Weinheim  (2006)  161--190.


\bibitem{Ha02}
B.~Hayes, {Follow the money}. {\em American Scientist} \textbf{90} (5),  (2002) 400--405.


\bibitem{Ioa}
Y. Ioannides and S. Skouras, US city size distribution: robustly Pareto, but only
in the tail. \emph{Journal of Urban Economics} \textbf{73}, (2013) 18--29.

\bibitem{IKR98}
S.~Ispolatov, P.L.~Krapivsky and S.~Redner, {Wealth distributions in
  asset exchange models}. {\em Eur. Phys. Jour. B} \textbf{2},   (1998)  267--276.
  
  \bibitem{KT}
D. Kahneman and A. Tversky. Prospect Theory: An Analysis of Decision under Risk,
\emph{Econometrica}, \textbf{47} (2)  (1979) 263--292.

\bibitem{KT1}
D. Kahneman, and A. Tversky. \emph{Choices, Values, and Frames}, Cambridge University Press, Cambridge 2000.
  

 \bibitem{Le09}
 M. Levy, Gibrat's law for (all) cities: comment. \emph{Am. Econ. Rev.}
\textbf{99}, (2009) 1672--1675.
 
\bibitem{LDD}
J. Luckstead, S. Devadoss and D. Danforth,
The size distributions of all Indian cities. \emph{Physica A} \textbf{474} (2017) 237--249.

\bibitem{LD1}
J. Luckstead and S. Devadoss,
 A comparison of city size distributions for China and India from 1950 to 2010.
\emph{Economics Letters} \textbf{124} (2) (2014) 290--295.
Jeff Luckstead, Stephen Devadoss

\bibitem{LD2}
J. Luckstead and S. Devadoss,
 Pareto tails and lognormal body of US cities size distribution.
\emph{Physica A} \textbf{465}  (2017)  573--578.

\bibitem{MD}
D. Maldarella and L. Pareschi, Kinetic models for socio--economic dynamics
of spe\-cu\-la\-tive markets, \emph{Physica A}, \textbf{391} (2012)  715--730 .

\bibitem{Mal}
Y. Malevergne, V. Pisarenko and D. Sornette,  Testing the Pareto against the
lognormal distributions with the uniformly most powerful unbiased test
applied to the distribution of cities. \emph{Phys. Rev. E} \textbf{83} (2011) 036111.

\bibitem{MZ}
M. Marsili and Yi-Cheng Zhang, Interacting Individuals Leading to Zipf's Law.
\emph{Phys. Rev. Lett.} \textbf{80}, (1998) 2741--2744.

\bibitem{MG}
A.C.R. Martins and S. Galam, Building up of individual inflexibility in opinion dynamics. \emph{Phys. Rev. E}, \textbf{87}  (2013) 042807.

 
\bibitem{MaTo07}
D.~Matthes and G.~Toscani,
\newblock On steady distributions of kinetic models of conservative economies.
\newblock {\em J. Stat. Phys.}  {\bf 130}  (2008)  1087--1117.

\bibitem{NPT}
G.~Naldi, L.~Pareschi and G.~Toscani,
\newblock {\em Mathematical modeling of collective behavior in socio-economic
  and life sciences}.
\newblock (Springer Verlag, Heidelberg, 2010).

\bibitem{New}
 M.E.J. Newman, Power laws, Pareto distributions and Zipf's law. \emph{Contemporary Phys.} \textbf{46}, (2005) 323--351.


\bibitem{PT14}
 L. Pareschi, G. Toscani, Wealth distribution and collective knowledge. A Boltzmann approach,  \emph{Phil. Trans. R. Soc. A} \textbf{372}, 20130396,  6 October (2014).

\bibitem{PT13}
L.~Pareschi and G.~Toscani,
\newblock {\em Interacting multiagent systems. Kinetic equations \& Monte Carlo
  methods}.
\newblock (Oxford University Press, Oxford, 2013).

\bibitem{Par}
V. Pareto, 
  {\it Cours d'{\'E}conomie Politique}.  Lausanne and Paris (1897).

\bibitem{PCG}
M. Patriarca, A. Chakraborti, G. Germano,
Influence of saving propensity on the power-law tail of the wealth distribution.
\emph{Physica A} \textbf{369}, (2006) 723--736.

\bibitem{PR}
M. Puente-Ajovín and A. Ramos,
On the parametric description of the French, German, Italian and Spanish city size distributions.
\emph{Ann. Reg. Sci.}  \textbf{54} (2015) 489--509.

\bibitem{Ram}
A. Ramos, 
Are the log-growth rates of city sizes distributed normally? Empirical evidence for the USA.
\emph{Empir. Econ.} \textbf{53}  (2017) 1109--1123.

\bibitem{PTo}
A. Pulvirenti, G. Toscani, Asymptotic properties of the inelastic Kac model, \emph{J. Statist. Phys.}, \textbf{114} (2004) 1453--1480.

\bibitem{Roz}
H. Rozenfeld, D. Rybski, X. Gabaix and H. Makse,  The area and population of
cities: new insights from a different perspective on cities. \emph{Am. Econ.
Rev.} \textbf{101}, (2011) 2205--2225.


\bibitem{SK13}
P. Sen and B. K. Chakrabarti, \emph{Sociophysics: An Introduction}. Oxford University Press, Oxford, 2013.


\bibitem{Sl04}
F.~Slanina, {Inelastically scattering particles and wealth distribution in an open eco\-no\-my}. {\em Phys. Rev. E} \textbf{69},  (2004)  046102.

\bibitem{SW}
K. Sznajd--Weron and J.  Sznajd, Opinion evolution in closed community.  \emph{Int. J. Mod. Phys. C} \textbf{11},
(2000)  1157--1165.


\bibitem{Tos06}
G.~Toscani,
\newblock Kinetic models of opinion formation.
\newblock {\em Commun. Math. Sci.}  {\bf 4}  (2006) 481--496.


\bibitem{TBD}
G. Toscani, C. Brugna and S. Demichelis, Kinetic models for the trading of goods. \emph{J. Stat. Phys.} \textbf{151}  549--566 (2013).


\bibitem{Vi}
C. Villani,
   \emph{Contribution {\`a} l'{\'e}tude math{\'e}matique des
  {\'e}quations de {B}oltzmann et de {L}andau en th{\'e}orie cin{\'e}tique des
  gaz et des plasmas}.
  {PhD Thesis, Univ. Paris-Dauphine}  (1998).

\bibitem{vil2}
 C. Villani,
 {On a new class of weak solutions to the spatially homogeneous Boltzmann
and Landau equations}.  {\em Arch. Rational Mech. Anal}, {\bf 143} (1998)
273--307.

\bibitem{Vil02} 
C. Villani, A Review of Mathematical Topics in
Collisional Kinetic Theory, in  S. Friedlander and D. Serre (eds),
{\em Handbook of Mathematical Fluid Dynamics}, Vol. 1, New York:
Elsevier (2002)

\bibitem{ZM}
D.H. Zanette and S.C. Manrubia, Role of intermittency in urban development: a model of large-scale city formation. \emph{Phys. Rev. Lett.} \textbf{79} (1997) 523--526.

\bibitem{Zipf}
 G.K. Zipf, Human behaviour and the principle of least effort. Addison--Wesley, Reading (MA) (1949).



\end{thebibliography}
\end{document}